# ACTIVATORS OF LUMINESCENCE IN SPELEOTHEMS AS SOURCE OF MAJOR MISTAKES IN INTERPRETATION OF LUMINESCENT PALEOCLIMATIC RECORDS

Y. Y. Shopov

**Abstract**
This work summarizes the main results of the operation of the International Program "Luminescence of Cave Minerals" of the commission on Physical Chemistry and Hydrogeology of Karst of UIS of UNESCO in the field of activators of speleothem luminescence. It discusses Activators of Luminescence in Speleothems as a source of major mistakes in the interpretation of luminescent paleoclimatic records. It demonstrates the existence of 6 types of luminescence of speleothems and cave minerals in dependence of the type of the luminescence center and its incorporation in the mineral. 24 different activators of photoluminescence of speleothem calcite and 11 of aragonite are studied. This paper demonstrates that it is impossible to produce reliable Paleotemperature or Past Precipitation records from luminescence of speleothems without establishing the organic origin of the entire luminescence of the particular sample.

**Introduction**
Absorption of excitation energy by a mineral leads to rising of electrons from ground state to an excited level. Sooner or later these electrons falls down to a lower level while emitting light. If the emission proceeds only during the excitation than it is called "fluorescence", if it proceeds later (usually seconds or minutes) than it is called "phosphorescence". In the later case falling of electrons from the excited state proceeds through intermediate levels (thus taking more time), so the energy of the emitted light is less than the energy of fluorescence (i.e. colour of the emitted light is shifted to the red). Some luminescent centers produce only fluorescence, but other both fluorescence and phosphorescence of minerals.
The type of luminescent centers determines the colour of luminescence. Colour may vary with changes of the excitation sources, because they may excite different luminescent centers existing in the mineral. Every luminescent center has its own excitation spectra Shopov, 1986), temperature dependence and conditions of excitation. One colour of luminescence sometimes may be produced by a single luminescent center or by a combination of two or several centers. The decay rate of luminescence (time for visible disappearance of the luminescence afterglow after switching off the excitation source) may vary from virtual zero for fluorescence to minutes or hours for phosphorescence. It is also characteristic for every luminescent center. Brilliance (brightness) of luminescence is function of the concentration of luminescence centers. It is almost linearly proportional to concentration of luminescent centers in transparent or white calcite, but can be substantially decreased by light absorption in colour centers of clay and other coloured inclusions or colour admixture ions in less-pure calcite.

Easiest and the most efficient method of excitation is irradiation by UV light sources producing photoluminescence and when luminescence is usually spoken about it is with this kind of excitation in mind. Phosphorescence of speleothems in caves can be seen by irradiating of speleothems with a photographic flash with closed eyes, with following rapid opening of the eyes after flashing. This simple technique is useful for the previous diagnostics of cave mineral and the selection of samples for laboratory analysis. Such "Visual Luminescent Analysis" (VLA) has been widely used in caves (TARCUS - CSSR, 1981), usually with a photographic flash but also with other simple devices such as portable UV lamps with short wave UV (SWUV) and long wave UV (LWUV). However data obtained by the VLA method are subjective and the determination of luminescence activators is not possible. In fact attempts to determine activators of the luminescence with VLA and chemical analysis leads to incorrect results.

It is known that almost 50 cave minerals have the capacity for exhibiting luminescence, but only 17 had been actually observed to be luminescent in speleothems so far (Shopov, 1997). This paper summarizes main results of the operation of the International Program "Luminescence of Cave Minerals" of the commission on Physical Chemistry and Hydrogeology of Karst of UIS of UNESCO in the field of activators of speleothem luminescence (Shopov 1989a).

**Origin of luminescence of Speleothems**

Many speleothems exhibit luminescence when exposed to ultraviolet (UV) or other light sources. In dependence of the type of the luminescence center and its incorporation in the mineral we distinguish following types of luminescence of speleothems and cave minerals:

*1. Luminescence of electron defects of the crystal lattice:*

Such type is the luminescence of $CO_3^{3-}$ ion in speleothem calcite under UV or electron beam excitation (Ugumory & Ikeya, 1980). It probably exists in any speleothem, but have lower quantum gain than the other types of luminescence in speleothems, so can be observed only in their absence. In cathodoluminescence petrography it is called "background luminescence". It is as intensive as older is calcite (Ugumory & Ikeya, 1980), because this center is produced only by ionising radiation from decomposition of natural radio- nuclides and have lifetime of millions of years. In ion crystals (such as chloride, fluorite or sulphide minerals) luminescence of this type is produced by admixtures of metal ions substituting the cations in the crystal lattice of the minerals. In this case the admixture cation must have different valency than the structural cations (Marfunin, 1979), so it cause compensation of the charge by trapping of free electrons or traps (which are activators of the luminescence of ion crystals).

*2. Luminescence of admixture ions substituting structural ions in the crystal lattice or incorporated in cavities of this lattice:*

Such type is the luminescence of most known luminescent centres in calcite, which are inorganic ions: $Mn^{2+}$, $Fe^{3+}$, $Pr^{3+}$, $Tb^{3+}$, $Er^{3+}$, $Dy^{3+}$, $Eu^{2+}$, $Eu^{3+}$, $Sm^{3+}$ and $Ce^{3+}$ (Tarashtan, 1978, Marfunin, 1979, Gorobets, 1981, Shopov, 1986, Shopov et al., 1988, Richter et al., 2003). This type of luminescence increases its intensity with decreasing of the temperature. This kind of luminescence exhibits strong quenching by $Fe^{2+}$, Ni and Cu- ions substituting structural cations in the crystal lattice, which adsorb the luminescence emission and re- emit it in the infrared region of the spectra (Marfunin, 1979).

### 3. Sensitizes luminescence of admixture ions substituting structural ions in the crystal lattice:

$Pb^{2+}$ have UV luminescence in calcite with no visible emission but it sensitizes the luminescence of $Mn^{2+}$, which produce short-time orange-red phosphorescence in hydrothermal calcites (Marfunin, 1979, Shopov, 1997). Such sensitized luminescence of these ions can be observed only if both they substitute a structural cation in the crystal lattice. $Mn^{2+}$ in calcite does not have strong absorption lines in UV, so it does not exhibit luminescence in infiltration calcites. $Pb^{2+}$ have very strong UV- absorption lines in calcite and transfer its excitation energy to $Mn^{2+}$ through the crystal lattice. It produces strong orange-red phosphorescence of $Mn^{2+}$ in calcite. This type of luminescence decreases its intensity with decreasing of the temperature, due to the reduction of the energy transfer through the temperature vibrations of the crystal lattice.

### 4. Luminescence of molecules, ions or radicals adsorbed inside of the lattice:

Such luminescence can be produced both by:

a. inorganic (like uranil ion- $UO_2^{2+}$) or

b. organic molecules (Tarashtan, 1978, Shopov, 1986, Shopov et al., 1988, White and Brennan, 1989, Shopov, 1997, 2002). In some cases they both produce luminescence of the same speleothem (fig.1).

This type of luminescence decreases its intensity with decreasing of the temperature, because energy transfer through the crystal lattice became impossible at low temperatures.

Usually luminescence of organics in speleothems is attributed to fulvic and humic acids (White and Brennan, 1989) but free acids could not exist in the alkaline karst environment. They react with the limestone producing their calcium salts in which form they exist in speleothems. The process of their chemical extraction from speleothems in order to study them converts them in free fulvic and humic acids.

Luminescence organics in speleothems can be divided to 4 types:- (1) Calcium salts of Fulvic acids, (2) Calcium salts of humic acids, (3) Calcium salts of huminomelanic acids (Shopov, 1997) and (4) Organic esters (Gilson et al., 1954). All these four types are usually present in a single speleothem with hundreds of chemical compounds with similar chemical behavior, but of different molecular weights. Concentration distribution of these compounds (and their luminescence spectra) depends on type of soils and plants over the cave, so the study of luminescent spectra of these organic compounds can give information about paleosoils and plants in the past (White, Brennan, 1989). Changes in visible colour of luminescence of speleothems suggesting major changes of plants society are observed very rare.

*5. Luminescence of inclusions of other minerals:*
Inclusions of other luminescent minerals can produce luminescence inside calcite speleothems. Most frequently these are inclusions of moon milk minerals. Such is also the green-yellow luminescence of magursilite clusters (Tarashtan, 1978) in speleothem calcite (Shopov, 1989b).

*6. Luminescence of fluid or gas inclusions*
Gas inclusions containing oil and gas products (hydrocarbons) had been observed to produce blue fluorescence and phosphorescence in speleothem calcites from Gaudalupe Mts., USA under SWUV or flash excitation (Shopov, 2001), but orange fluorescence under LWUV excitation.

All six types of luminescence centers are observed to produce luminescence of speleothem calcites under UV excitation.
Different types of excitation may excite different luminescent centers. Some or all of them may produce luminescence in a single speleothem (Shopov, 1997, 2001, Richter et al., 2003).

**Activators of Luminescence as Source of Mistakes in Interpretation of Luminescent Paleoclimatic Records**
Recently some researchers attribute all luminescence in calcite speleothems to organics (e.g. Baker et al, 1993) without any reason to do so. But 14 (58% of all known) activators of speleothem luminescence are inorganic. Minerals are not pure chemical substances and contain many admixtures. Usually several centres activate luminescence of one sample (table 1) and the measured spectrum is a sum of the spectra of two or more of them (fig.1). Luminescence of minerals formed at normal cave temperatures (below 40º C) is usually (but far not always) due mainly to molecular ions and absorbed organic molecules. Luminescence of uranil- ion ($UO_2^{2+}$) is also very common (fig.1) in such speleothems (Shopov, 2001). Luminescence of other inorganic ions sometimes dominate luminescence spectrum of speleothems.
Before 1983 all luminescence in calcite speleothems was attributed to inorganic ions (Kropachev et al. 1971, Mitsaki, 1973, Slacik, 1977, Turnbull, 1977, Ugumory & Ikeya, 1980, Rogers and Williams 1982, Hill and Forti, 1986)
All paleoenvironmental luminescence (paleoluminescence) methods (Shopov, 2004) use only luminescence of organics in Speleothems. Therefore it is necessary to determine that all luminescence of the sample is due to organics before using a speleothem for any paleo environmental work. Detailed spectral measurements of the luminescence are absolutely necessary to determine luminescent compounds in any speleothem. This requires the use of a luminescence spectrometer, plus an Electron Spin Resonance (ESR) spectrometer or chromatograph (Shopov, 1989a,b). Lasers and Raman spectrometers used for measurements of luminescent spectra allow also determination of the luminescent mineral or inclusion in the speleothem, because the narrow Raman lines appearing in luminescence spectra at high resolution scanning are characteristic for different minerals.
In many calcite speleothems all or a significant part of the luminescence is produced by inorganic ions (Shopov, 1986, Shopov et al., 1988). Sometimes they even have annual banding (photo 1) due to variations of acidity of the karst waters, causing variations of the solubility of some inorganic luminophores (Shopov, 1997). Uranium compounds have such migration behavior. We found some speleothems demonstrating fine fluorescence banding produced by uranium impurities in the speleothem (photo 1) under shortwave UV light (Shopov, 2002). Fine fluorescence banding under long-wave UV light is produced by rare earth elements in the same sample. This banding can be annual or even sub- annual. Such luminescence banding is usually considered to be annual (if produced by organics) and have a number of paleoclimatic and dating applications (Shopov et al., 1997). Phosphorescence of this sample (not shown) suggests that there are no any luminescent organics in the middle (darker) part of the speleothem, but there are some in the outer part of the sample.

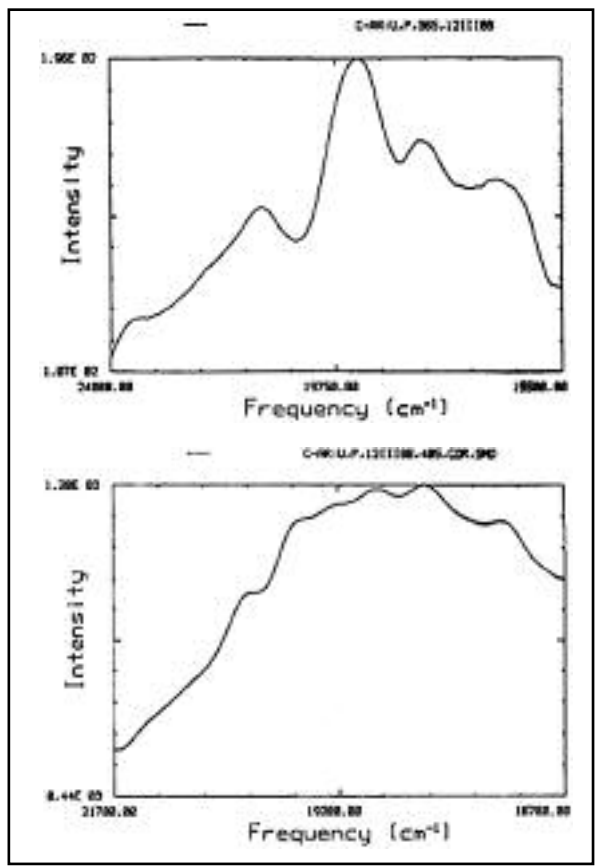

Figure 1. Luminescence of speleothem calcite under excitation by 365 nm (up) and 405 nm (down) lines of Hg- lamp. The narrow lines of luminescence in both spectra are produced by uranil- ion ($UO_2^{2+}$) while the broadband luminescence is due to organics. UV excitation of $UO_2^{2+}$ is far more efficient than this of organics so it predominates in the spectra at 365 nm excitation.

Statements that Sr causes violet luminescence of carbonate speleothems (e.g. Kropachev et al. 1971), Zn greenish- white luminescence of calcite stalactites (Turnbull, 1977) and Cu- causes pale-green and blue luminescence of calcite and aragonite (e.g. Rogers and Williams 1982) are in error. Sr and Zn- ions do not have electron transitions in the visi-

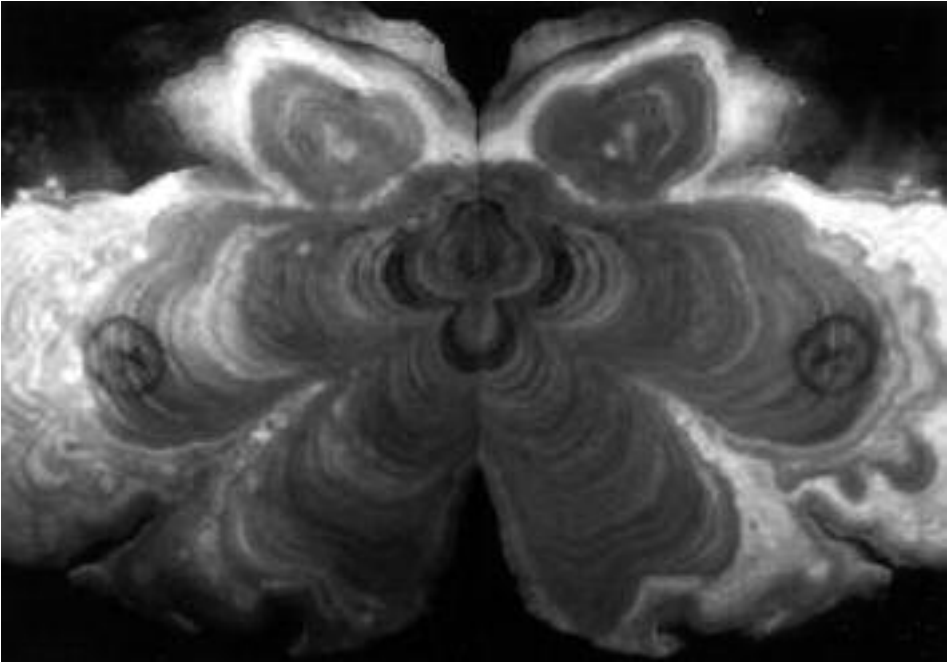

Foto 1. Annual (?) banding of luminescence of uranil- ion and rare earth elements in a calcite flowstone due to variations in pH of the water (Luminescence of a section of a calcite coralloid from a cave near Irkutsk, Russia). Fine fluorescence banding under short-wave UV light (left) is produced by uranium impurities ($UO_2^{2+}$) in the speleothem. Fine fluorescence banding under long-wave UV light (right) is produced by rare earth elements in the same sample (right and left images are mirror images of the same sample). Photo by Yavor Shopov.

ble region of the spectra and therefore cannot activate luminescence in carbonates, but Cu is known to cause quenching (reduction of luminescence) induced by other cations (Tarashtan 1978). $Cu^{2+}$ can excite only infrared luminescence of some sulfides. Also, interpretations of the visible luminescence of calcite as Pb- activated (Slacik, 1977) are not correct, because Pb in calcite emits only UV light (Tarashtan 1978, Shopov et al. 1988). Such wrong interpretations had been obtained by correlation of the intensity of luminescence with the concentration of these elements in speleothems without proper measurements of spectra of their luminescence.

Luminescence of the high- temperature hydrothermal minerals is due mainly to cations because molecular ions and molecules destruct at high temperatures. The orange-red luminescence of $Mn^{2+}$ in calcite (table 1) sensitized by $Pb^{2+}$ can be observed only in hydrothermal calcite, because $Pb^{2+}$ has very big ion radius and can substitute $Ca^{2+}$ in the crystal lattice of calcite only at high temperatures, so it can be used as an indicator of the hydrothermal origin of the cave mineral (Shopov, 1989 a, b). Therefore, if calcite has only orange-red, short time phosphorescence, it is sure to have formed in high-temperature, hydrothermal solutions (>300º C). But if it has long-time phosphorescence in addition to the red- orange one, then it is a low-temperature hydrothermal calcite (Shopov,

## Table 1. Activators of Luminescence of Speleothems

| Luminescence activator | Excitation | Emission color | Afterglow | Origin | Reference |
|---|---|---|---|---|---|
| **Calcite:** | | | | | |
| 1. Esters | Hg-lamp | blue | long | infiltration | Gilson et al. (1954) |
| 2. Organics, | $N_2$-Laser | blue | long | infiltration | Shopov et al. (1983) |
| 3. Calcium salts of Fulvic & humic acids | Ar-L., Xe | yellow-green | long | infiltration | Shopov et al. (1989) |
| 4. Fulvic & humic acids | SWUV | blue-green | long | infiltration | White, Brennan (1989) |
| 5. Organics | Ar-L, Xe -lamp | blue-green | long | infiltration | Shopov (1989b) |
| 6. Organics | $N_2$-Laser | yellow-green | long | infiltration | Shopov (1989b) |
| 7. Organics | LWUV, Hg | yellow | long | infiltration | Shopov (1989b) |
| 8. Organics | SWUV, LWUV | yellow-orange | long | infiltration | White, Brennan (1989) |
| 9. $CO_3^{3-}$ | $N_2$-Laser | blue | | infiltration | Ugumory, Ikeya (1980) |
| 10. $UO_2^{2+}$ | SWUV | green | no | infiltration | White, Brennan (1989) |
| 11. $UO_2^{2+}$ | $N_2$-L., Hg -lamp | green | no | infiltration | Shopov (1989b) |
| 12. *$UO_2^{2+}$(magursilite?) | $N_2$-L., Hg -lamp | green- yellow | no | infiltration | Shopov (1989b) |
| 13. Organics | Hg, Xe -lamp | bluish | <15s | hydrothermal | Dublyansky (in press) |
| 14. *$Mn^{2+}$ | Ar-L, $N_2$-L., Xe | orange-red | 0.1s | h-t.hydrothermal | Mitsaki,1973;White,1974 |
| 15. *Hydrocarbons | Xe –flash lamp | violet | long | epithermal | Shopov et al (1996) |
| 16. $Fe^{3+}$ | Ar- Laser | dark- red | | ? hydrothermal | Shopov, 1988 |
| 17. $Pb^{2+}$ | SWUV | UV | | hydrothermal | Shopov, 1985 |
| 18-24*. Rare Earth Elements $^{3+}$-ions | LWUV, SWUV electrons | various | | ? | Shopov, 1985, 1988 Richter (2002) |
| **Aragonite:** | | | | | |
| 25. Organics | Hg-lamp | blue | long | infiltration | Shopov (1989b) |
| 26. Organics | Hg-lamp | blue-green | long | infiltration | Shopov (1989b) |
| 27. Organics | $N_2$-Laser | blue-green | long | infiltration | Shopov (1989b) |
| 28. Organics | $N_2$-Laser | green | long | infiltration | Shopov (1989b) |
| 29. Organics | Hg-lamp | yellow | long | infiltration | Shopov (1989b) |
| 30. Organics | SWUV, LWUV | blue-green | long | infiltration | White, Brennan (1989) |
| 31. $UO_2^{2+}$ | SWUV | green | no | infiltration | White, Brennan (1989) |
| 32. ? | Hg (LWUV) | orange | ? | ? | White, Brennan (1989) |
| 33. $Mn^{2+}$ | LWUV, e-beam | yellow-green | short | ? | Shopov, 1988 |
| 34. $Sm^{3+}$ | LWUV, e-beam | red | | ? | Shopov, 1985 |
| 35. $Eu^{2+}$ | LWUV, e-beam | blue | | ? | Shopov, 1988 |

**Comments to table 1:**
* - luminescence of Rare Earth elements in calcite is well described in (Tarashtan, 1978, Shopov, 1986, Shopov et al., 1988, Richter, 2002), so is not included in the table
*12- Tarashtan (1978) attributed this spectrum of luminescence to luminescence of clusters of the mineral magursilite adsorbed in calcite;
*14- also in (Shopov et al., 1988, White and Brennan 1989)
*15 - hydrocarbons present only in fluid inclusions in calcite, formed 1 km below the surface by waters heated by Earth thermal gradient (epithermal solutions) in a cave in Carlsbad Caverns region. Guadeloupe Mts., New Mexico, US (Shopov et al., 1996)

1989a,b). Calcites formed by low-temperature hydrothermal solutions have fluorescence or short-life phosphorescence due to cations and long phosphorescence due to molecular ions (Gorobets, 1981). Minimal temperature of appearance of this orange- red luminescence was estimated to be of about 40$^0$C by Dublyansky (in press) by fluid inclusion analysis in hydrothermal cave calcites, but our direct measurements of luminescence of calcites in hot springs shows that even at 46$^0$C such luminescence do not appear (Petrusenko et al., 1999). It probably appears at over 60$^0$C. Luminescence of hydrothermal calcite formed at lower temperatures looks similar to usual speleothem luminescence. Such luminescence data visualize the changes of the temperature of mineral forming solutions and are comparable with the stable isotope data used conventionally for this purpose (Bakalowicz et. al., 1987, Ford et al., 1993).

**Conclusions**
Before using of any speleothem for paleoenvironmental luminescence measurements it is necessary to determine that all luminescence of the sample is due to organics. Otherwise interpretation of the data can be completely wrong and there is no way to prove or disapprove it without further measurements on the same sample to establish the organic nature of all its luminescence.

**Acknowledgements**
This research was funded by Bulgarian Science Foundation by research grant 811/98 to Y. Shopov

**References**

Bakalovicz M., Ford D.C., Miller T.E., Palmer A.N., Palmer M..V., 1987 - *Thermal genesis of dissolution cave from Black Hills, South Dakota*. GSA Bulletin, 99: 729-738.
Baker A., Smart P.L., Edwards R.L., Richards D.A, 1993 - *Annual Growth banding in a cave stalagmite*. Nature, 304: 518-520.
Dublyansky Y.V., 2004 - *Luminescence of calcite from Buda Hill caves*. (in press)
Ford D.C., Bakalovicz M., Miller T.E., Palmer A.N., Palmer M..V. 1987 - *Uranium series dating of the draining of an aquifer: an example of Wind cave Black Hills, South Dakota*. GSA Bulletin, 105: 241-250.
Gilson R.J., MCarthney E., 1954 - *Luminescence of speleothems from Devon, U.K.: The presence of organic activators*. Ashford Speleological Society Journal, 6: 8.(abstr.)
Gorobets B.S., 1981 - *Atlas of Spectra of Luminescence of Minerals*. VIMS, Moskow. (in Russian)
Hill C., Forti P., 1986 - *Cave Minerals of the World*. NSS, Huntsville, Alabama, USA: 238 p.
Kropachev A.M., Gorbunova K.A., Tzikin V.Y., 1971 - *Metachromatism and luminescence of the carbonate speleothems from the caves of Bashkiriya form Krasnoyarks region*. Peshtery, 10-11:74-80. (in Russian)
Marfunin A. S., 1979 - *Spectroscopy Luminescence and Radiation Centers in Minerals*. Berlin, Springer-Verlag: 352 p.
Mitsaki V., 1973 - *Geochemical study of some specimens of stalactites from Tourkovonia cave, Athens*. Deltio, 12(3): 90-95.
Petrusenko S., Shopov Y.Y., Kunov A., 1999 - *Micromorphologic Peculiarities and Typomorphic*

*Luminescence of Calcites from Bulgarian Deposits of Various Genesis*. Proceedings of The National Scientific Conference on New Achievements and Actual Problems of Karstology and Speleology in Bulgaria, March 25- 28, 1999, Sofia: 75- 81.

Richter D.K., Goette Th., Niggemann S., Wurth G., 2003 - *Cathodoluminescence of Carbonate Speleothems: State of the Art*. In: Carrasco F., Duran J.J., Andreo B. (Eds.): *Karst and Environment*: 381- 387.

Rogers B.W., Williams K.M., 1982 - *Mineralogy of Lilburn cave, Kings Cannion National Park, California*. NSS- Bulletin, 44(2): 23-31

Shopov Y.Y., Spasov V.A., 1983 - *Speleological Applications of Physical Methods for analysis of solids*. Abstracts of the First National Congress of the Physicists in Bulgaria (Sofia, 28 IX- 1 X 1983): 193.

Shopov Y.Y., 1986 - *Applications of photoluminescence in Speleology*. Bulgarian Caves, 4: 38-45.

Shopov Y.Y., 1989a - *Bases and Structure of the International Programme "Luminescence of Cave Minerals" of the Commission of Physical Chemistry and Hydrogeology of Karst of UIS*. Expedition Annual of Sofia University, 3/4: 111- 127.

Shopov Y.Y., 1989b - *Spectra of Luminescence of Cave Minerals*. Expedition Annual of Sofia University, 3/4: 80-85.

Shopov Y.Y., 1997 - *Luminescence of Cave Minerals*. In: Hill C., Forti P. (Eds.): *Cave Minerals of the world*, second edition, NSS, Huntsville, Alabama, USA: 244-248.

Shopov Y.Y., Ivanov G.I., Kostov R.I., 1988 - *Phase Analysis of the Polymorphic modifications of $CaCO_3$ by spectra of its photo- luminescence*. In: "Spectral Methods of solving of the problems of the Solid State Physics", Scientific Commission on "Atomic and Molecular Spectroscopy" of AS USSR, Acad. Sci. Press, Moskow: 191-201 (in Russ.)

Shopov Y.Y., Tsankov L., Buck M., Ford D.C., 1996 - *Time Resolved Photography of Phosphorescence- A New Technique for Study of Thermal History and Uplift of Thermal Caves*. Extended abstracts of Int. Conference on "Climatic Change- the Karst Record", 1- 4 August 1996, Bergen, Norway. Karst Waters Institute Special Publication 2: 154.

Slacik J., 1977 - *Luminescence analysis in speleology*. Proc. 7[th] Int. Cong. Speleol. Sheffield: 31-36.

Tarashtan A.N., 1978 - *Luminescence of minerals*. Naukova Dumka, Kiev (in Russian)

TARCUS-CSSR, 1981 - *Bibliography of working group of the CSS ZO 1-O5 GEOSPELEOS for 1973-1980*. Prague ,CSS, 1-31.

Turnbull, I.C., 1977 – Zinc: an activator of fluorescence in cave calcite. Fluorescent Min. Soc. Journ., 6: 58-60.

Ugumori T., Ikeya M., 1980 - *Luminescence of $CaCO_3$ under $N_2$- Laser excitation and application to archaeological dating*. Japanese J. Appl. Physics, 19(3): 459-465.

White W.B., 1974 - *Determination of Speleothem growth mechanism by luminescence spectrography*. Geo2, 3(3): 37.

White W.B, Brennan E.S., 1989 - *Luminescence of speleothems due to fulvic acid and other activators*. Proceedings of 10th International Congress of Speleology, 13-20 August 1989, Budapest, 1: 212-214.